\documentclass[conference,final]{IEEEtran}

\usepackage[final]{graphicx}
\usepackage[reqno]{amsmath}
\usepackage{amssymb}

\newfont{\bbb}{msbm10 scaled 500}

\newfont{\bb}{msbm10 scaled 1100}

\newcommand{\Prob}{\textrm{Pr}}

\newcommand{\hv}{{\bf h}}

\newcommand{\Wc}{{\cal W}}

\newcommand{\Xc}{{\cal X}}
\newcommand{\Yc}{{\cal Y}}

\newtheorem{theorem}{Theorem}
\newtheorem{lemma}[theorem]{Lemma}

\IEEEoverridecommandlockouts

\title{On the Delay Limited Secrecy Capacity of Fading Channels}

\author
{
\IEEEauthorblockN{Karim Khalil and Moustafa Youssef}
\IEEEauthorblockA{
Wireless Intelligent Networks Center (WINC) \\
Nile University, Cairo, Egypt \\
Email: \{kareem.makarem,mayoussef\}@nileu.edu.eg}
\and
\IEEEauthorblockN{O. Ozan Koyluoglu and Hesham El Gamal}
\IEEEauthorblockA{
Department of Electrical and Computer Engineering \\
Ohio State University, Columbus, Ohio \\
Email: \{koyluogo,helgamal\}@ece.osu.edu}
\thanks{
Hesham El Gamal also serves as the Director for the
Wireless Intelligent Networks Center (WINC), Nile University, Cairo,
Egypt.
This work is supported in part by an NPRP grant from
the QNRF and the Egyptian NTRA.
The statements made herein are solely the responsibility
of the authors.
}
}

\begin{document}
\maketitle


\begin{abstract}

In this paper, the delay limited secrecy capacity of the flat fading
channel is investigated under two different assumptions on the available
transmitter channel state information (CSI). The first scenario assumes
perfect prior knowledge of both the main and eavesdropper channel gains.
Here, upper and lower bounds on the secure delay limited capacity
are derived and shown to be tight in the high signal-to-noise ratio
(SNR) regime (for a wide class of channel distributions). In the second
scenario, only the main channel CSI is assumed to be available at the
transmitter. Remarkably, under this assumption, we establish the
achievability of non-zero secure rate (for a wide class of channel
distributions) under a strict delay constraint. In the two cases,
our achievability arguments are based on a novel two-stage approach
that overcomes the {\em secrecy outage} phenomenon observed in
earlier works.

\end{abstract}


\section{Introduction}

In many applications, there is a delay constraint on the
data to be transmitted via a wireless link. These applications
range from the most basic voice communication to the more demanding
multimedia streaming. However, due to its broadcast nature, the
wireless channel is vulnerable to eavesdropping and other security
threats. Therefore, it is of critical importance to find techniques
to combat these security attacks while satisfying the delay limitation
imposed by the Quality of Service (QoS) constraints. This motivates our
analysis of the fundamental information theoretic limits of secure
transmission over fading channels subject to strict deadlines.

Recent works on information theoretic security have been largely
motivated by Wyner's wire-tap channel model~\cite{Wyner:The75}.
In his seminal work, Wyner proved the achievability of non-zero
secrecy capacity, assuming that the wiretapper channel is a
degraded version of the main one, by exploiting the noise to
create an advantage for the legitimate receiver. The effect
of fading on the secrecy capacity was further studied in~\cite{Gopala:On08}
in the ergodic setting. The main insight offered by this work is that
one can {\em opportunistically} exploit the fading to achieve
a non-zero secrecy capacity even if the eavesdropper channel
is better than the legitimate receiver channel, on the average.

Delay limited transmission over fading channels has been well studied
in different network settings and using various traffic models. For example,
in~\cite{Hanly:Multiaccess:98}, the delay limited capacity notion was
introduced and the optimal power control policies were characterized in
several interesting scenarios. In~\cite{Berry:Communication02}, the strict
delay limitation of~\cite{Hanly:Multiaccess:98} was relaxed by allowing for
buffering the packets at the transmitter. In this setup, the asymptotic
behavior of the power-delay trade-off curve was characterized yielding
valuable insights on the structure of the optimal resource allocation
strategies~\cite{Berry:Communication02}. More recently, the scheduling
problem of data transmission over a finite delay horizon assuming
perfect CSI was considered~in~\cite{Lee:Energy-efficient}. Our work can
be viewed as a generalization of~\cite{Hanly:Multiaccess:98} where a
secrecy constraint is imposed on the problem. The extension to the bursty
traffic scenario is currently under investigation.

The delay limited transmission of secure data over fading channels
was considered previously in~\cite{Bloch:Wireless08}. In this work,
the authors attempted to send the secure information using binning
techniques inspired by the wiretap channel results. The drawback
of this approach is that it fails to secure the information in
the particular instants where the eavesdropper channel gain is
larger than that of the main channel resulting in the so-called
{\bf secrecy outage} phenomenon (as defined in \cite{Bloch:Wireless08}).
Unfortunately, in the delay limited setting, the secrecy outage
can not made to vanish by increasing the block length
leading to the conclusion that the delay limited rate
achieved by this approach is equal to zero for most
channel distributions of interest~\cite{Bloch:Wireless08}.
This obstacle is overcome by our two-stage approach. Here, the
delay sensitive data of the current block is secured via
Vernam's one time pad approach~\cite{Vernam:OneTimePad26},
which was proved to achieve perfect secrecy by
Shannon~\cite{Shannon:Communication49}, where the legitimate nodes
agree on the private key during the previous blocks.
Since the key packets are {\bf not delay sensitive}, the two nodes
can share the key by distributing its bits over many fading
realizations to capitalize on the ergodic behavior of the channel.
Through the appropriate rate allocation, the key bits can be
{\bf superimposed} on the delay sensitive data packets so that
they can be used for securing future packets. This is referred as
key renewal process in the sequel. This process requires an
\emph{initialization phase} to share the key needed for securing the
first data packets. However, the loss in the secrecy entailed by
the initialization overhead vanishes in the asymptotic limit of
a large number of data packets. Our analytical results establish
the asymptotic optimality, with high SNR, of this novel approach
in the scenario where both the main and eavesdropper channel
gains are known {\em a-priori} at the transmitter (for a wide
class of channel distributions). When only the main channel
CSI is available, this approach is shown to achieve
{\bf a non-zero constant} secrecy rate for a wide class of
{\em quasi-static channels} (i.e., the class of invertible
channels~\cite{Hanly:Multiaccess:98}).

The rest of the paper is organized as follows. Section II
details the system model and the notations used throughout the
rest of the paper. In Section III, our main results for both the
full and main CSI cases are obtained. Finally, Section
IV concludes the paper.


\section{System Model}

\begin{figure}[t] 
    \centering
    \includegraphics[width=0.9\columnwidth]{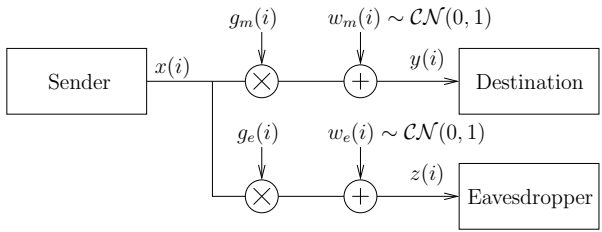}
    \caption{System Model}
    \label{model}
\end{figure}

The system model is as shown in Figure~\ref{model}. A source node
(Alice) communicates with a destination node (Bob) over a fading
channel in the presence of an eavesdropper (Eve). We adopt a block
fading model, in which the channel is assumed to be constant during a
coherence interval and changes randomly from an interval to another
according to a bounded continuous distribution. Also the coherent
intervals are assumed to be large enough to allow for the use of
random coding techniques.

During any coherence symbol interval $i$, the signals received at the
destination and the eavesdropper are given by
\begin{eqnarray}
y(i)&=&g_m(i)\, x(i)+w_m(i),\\
z(i)&=&g_e(i) x(i)+w_e(i),
\end{eqnarray}
where $x(i)$ is the transmitted symbol, $g_m(i)$ and $g_e(i)$ are
the main channel and the eavesdropper channel gains
respectively, $w_m(i)$ and $w_e(i)$ are the i.i.d. additive white
complex gaussian noise with unit variance at the legitimate receiver
and the eavesdropper, respectively. We denote the power gains
of the fading channels for the main and eavesdropper channels
by $h_m(i) = |g_m(i)|^2$ and $h_e(i) = |g_e(i)|^2$, respectively.
We impose the long term average power constraint $\bar{P}$, i.e.,
\begin{eqnarray}
{\mathbb E} [P(\hv)]\leq \bar{P},
\end{eqnarray}
where $P(\hv)$ is the power allocated for the channel state
$\hv=(h_m,h_e)$ and the expectation is over the channel gains.

The source wishes to send a message $W \in \Wc =\{1,2,\cdots ,M\}$
to the destination. {\bf In the following, our delay
constraint is imposed by breaking our message
into packets of equal sizes, where each one is encoded
independently, transmitted in only one coherence block,
and decoded by the main receiver at the end of this block}.
Accordingly, the total number of channel uses is partitioned
into $b$ super-blocks. Each super-block is divided into
$a$ blocks of $n_1$ symbols, where $n=b\ a\ n_1$ and
$n_1$ denotes the length of coherence intervals.
We will further represent a fading block with tuple
$(m,l)$ such that $m \in \{1,2,\cdots,b\}$ and
$l \in \{1,2,\cdots,a\}$. We consider the problem of
constructing $(M_1,n_1)$ codes ($M=b\ a\ M_1$) to transmit
the message of the block $(m,l)$,
$W(m,l)\in \Wc_1 = \{1,2, \dots, M_1\}$ to the receiver.
Here, an $(M_1,n_1)$ code consists of the following
elements: 1) a stochastic encoder $f_{n_1}(.)$ at the
source that maps the message $w(m,l)$ to a codeword
$x^{n_1}(m,l) \in \Xc ^{n_1}$, and 2) a decoding
function $\phi$: $\Yc^{n^*} \rightarrow \Wc_1 $
at the legitimate receiver, where $n^*= (m-1)an_1+ln_1$
denotes the total number received signal dimension at
the receiver at the end of the block $(m,l)$. The average error
probability of an $(M_1,n_1)$ code is defined as
\begin{eqnarray}
P_e^{n_1}=\frac{1}{M_1}
\sum \limits_{w \in \Wc_1}
\Prob (\phi (y^{n^*})\neq w
| w \mbox{  was sent}), \nonumber
\end{eqnarray}
where $y^{n^*}$ represents the total received signals at the
legitimate receiver at the end of the block $(m,l)$.
The equivocation rate $R_e$ at the eavesdropper is
defined as the entropy rate of the transmitted message
conditioned on the available CSI and the channel outputs
at the eavesdropper, i.e.,
\begin{eqnarray}
R_e \overset{\Delta}{=}
\frac{1}{n} H(W|Z^n,h_m^n,h_e^n),
\end{eqnarray}
where $h_m^n=\{ h_m(1), \cdots, h_m(n)\}$
and $h_e^n = \{ h_e(1), \cdots, h_e(n)\}$ denote
the channel power gains of the legitimate receiver and the
eavesdropper in $n$ symbol intervals, respectively. We consider
only the perfect secrecy (in the sense of~\cite{Wyner:The75})
which requires the equivocation rate $R_e$ to be $\epsilon$
close to the message rate.
The delay limited perfect secrecy rate $R_{s,d}$ is said to
be achievable if for any $\epsilon>0$, there exists a
sequence of codes $(2^{n_1R_{s,d}},n_1)$
such that for any $n_1\geq n_1(\epsilon)$, we have
\begin{eqnarray}
P_e^{n_1} &\leq& \epsilon \nonumber \\
R_e &\geq& R_{s,d}-\epsilon \nonumber
\end{eqnarray}
for any fading block $(m,l)$.

Finally, we give some notational remarks.
We denote the delay limited secrecy rate and capacity as
$R_{s,d}^{(F)}$ and $C_{s,d}^{(F)}$ for the full CSI case
(both $g_m$ and $g_e$ are known {\em a-priori} at the transmitter).
We respectively use the notation $R_{s,d}^{(M)}$ and $C_{s,d}^{(M)}$
for the main CSI case (only $g_m$ is known {\em a-priori} at the transmitter).
We denote $[x]^+ = \max\{x,0\}$. And, we remark that the expectations
are with respect to channel gains throughout the sequel.


\section{Main Results}

\subsection{Full CSI Scenario}
First, we give a simple upper bound on the delay limited secrecy
capacity that will be used later to establish the optimality of the
proposed approach in the high SNR regime.

\begin{theorem}\label{thm:upperFullCSI}

The delay-limited secrecy capacity when both $g_m$ and $g_e$
are available at the transmitter, $C_{s,d}^{(F)}$, is upper bounded by
\begin{eqnarray}\label{eq:upperfull}
C_{s,d}^{(F)} \leq
\max \limits_{{}^{\quad \: P(\hv)}_{\textrm{s.t. }
{\mathbb E} [P(\hv)]\leq \bar{P} }}
\min
\left\{  R_s^{(F)},  R_d^{(F)}  \right\},
\end{eqnarray}
where $R_s^{(F)}$ and  $R_d^{(F)}$ are given as follows.
\begin{eqnarray}
R_s^{(F)} &=& \mathbb{E} \left[\log(1+P(\hv)h_m)-
\log(1+P(\hv)h_e)\right]^+ \nonumber \\
R_d^{(F)} &=& \min\limits_{\hv} \log(1 + P(\hv)h_m ) \nonumber
\end{eqnarray}
\end{theorem}

\begin{proof}
For a given power allocation scheme $P(\hv)$,
we have
\begin{eqnarray}\label{eq:thm1eq1}
R_{s,d} ^{(F)} \leq R_s^{(F)},
\end{eqnarray}
for any achievable delay limited secrecy rate
$R_{s,d}^{(F)}$, since imposing delay constraint
can only degrade the performance. We also have,
for a given $P(\hv)$,
\begin{eqnarray}\label{eq:thm1eq2}
R_{s,d} ^{(F)} \leq R_d^{(F)},
\end{eqnarray}
since imposing secrecy constraint can not
increase the achievable rate. Then,
combining \eqref{eq:thm1eq1} and \eqref{eq:thm1eq2},
and maximizing over $P(\hv)$, we obtain
\begin{eqnarray}\label{eq:thm1eq3}
R_{s,d} ^{(F)} \leq \max\limits_{P(\hv)}
\min \{ R_d^{(F)}, R_s^{(F)}\},
\end{eqnarray}
for any achievable delay-limited secrecy rate
$R_{s,d} ^{(F)}$, which proves our result.
\end{proof}

The following result establishes a lower bound on the delay
limited secrecy capacity using our novel two-stage approach. The
key idea is to share a private key between Alice and Bob, without
being constrained by the delay limitation. Then, this key is used
to secure the delay sensitive data to overcome the secrecy outage
phenomenon. In the steady state, the key renewal process takes place
by superimposing the key on the delay sensitive traffic. More
precisely, as outlined in the proof, the delay sensitive traffic
(secured by the previous key) serves as a {\em randomization} signal
in the binning scheme used to secure the current key. Finally,
since $h_e$ is known {\em a-priori} at the transmitter, one
can further increase the delay limited secrecy rate by dedicating
a portion of the secure rate to the delay sensitive traffic
(as controlled by the function $q(\hv)$ in the following theorem).

\begin{theorem}\label{thm:lowerFullCSI}

The delay-limited secrecy capacity in the full CSI
scenario is lower bounded as follows.
\begin{eqnarray}
C_{s,d} ^{(F)} \geq R_{s,d}^{(F)}
= \max \limits_{{}^{P(\hv), \: q(\hv)}_{\textrm{s.t. }
{\mathbb E} [P(\hv)]\leq \bar{P} }}
\bigg[
\min \limits_{\hv}
\left\{
R_s^{\prime\prime}(\hv) + R_o(\hv)
\right\}
\bigg],
\end{eqnarray}
where
\begin{eqnarray}
&R_s^{\prime\prime}(\hv)
= R_s(\hv) - R_s^{\prime}(\hv), \nonumber\\
&R_s(\hv) = \left[
\log(1+P(\hv)h_m)- \log(1+P(\hv)h_e)
\right]^+, \nonumber\\
&R_s^\prime(\hv) =  [\log(1+P(\hv)h_m)-
\log(1+P(\hv)q(\hv))]^+, \nonumber
\end{eqnarray}
such that
$q(\hv)$ is an arbitrary chosen function
satisfying $q(\hv)~\geq~h_e\:$  $\forall~h_e$, and
$R_o$ is chosen to satisfy the followings.
\begin{eqnarray}\label{eq:thm2eq3}
&\mathbb{E}[R_o(\hv)] \leq \mathbb{E}[R_s^\prime(\hv)]\nonumber\\
&R_o(\hv) \leq \min \left\{\log(1+P(\hv)h_m),\log(1+P(\hv)h_e)\right\}
\end{eqnarray}

\end{theorem}
\begin{proof}[Sketch of the Proof]
In our scheme, we require Alice to transmit a delay constrained
data message and a key to Bob. The key is used to encrypt
data and thus should be secured from Eve.  A given message
$w\in\{1,2,\cdots, 2^{(n(R_{s,d}^{(F)}))}\}$
is transmitted by sending $ba$ data packets of equal length,
each represented by $D(m,l)$,
where each packet is {\bf encoded independently}
and sent with rate $R_{s,d}^{(F)}$ during the
corresponding block of the channel. We further divide a packet
to be transmitted at block $(m,l)$ into two parts $D_1(m,l)$ and
$D_2(m,l)$. The first part of data packet is transmitted along
with the generated key using the one-time pad scheme, whereas
the second part is transmitted as a secret message. We use a
separation strategy similar to~\cite{Prabhakaran:Secrecy08}
by sending public and private messages simultaneously. But,
in contrast to~\cite{Prabhakaran:Secrecy08}, we here have
the fading channel as the resource from Alice to Bob and
Eve and we exploit it to secure the key, and hence, the message.
We now describe the initial key generation and key renewal.
For the very first $a$ blocks (the super-block $m=1$), we
transmit the key, $K(1)$, from Alice to Bob securely. Utilizing
the ergodicity of the channel, we can transmit a key of length
$an_1 \mathbb{E}[R_s'(\hv)]$ bits~\cite{Gopala:On08}.
Then, for any super-block $m$,
we will use the key $K(m-1)$ for the one time pad, and also
generate a new key $K(m)$ for the use in the next super-block.
For any given block $(m,l)$, we use the $n_1R_o(\hv)$
remaining bits of the key $K(m-1)$ and denote the
corresponding bits as $\tilde{K}(m,l)$.
These bits are used in a one-time pad scheme to construct
$D_o(m,l) = D_1(m,l) \oplus \tilde{K}(m,l)$. The encrypted bits
are then mapped to a message $w_1(m,l)\in\{1,2,\cdots,2^{n_1R_o(\hv)}\}$.
The message $w_1$ along with a possible
additional randomization is transmitted along with the secret
data. Here, the secret data we sent within a block is two-fold:
$w_2(m,l)\in\{1,2,\cdots,2^{(n_1R_s''(\hv))}\}$ which carries
the corresponding data $D_2(m,l)$ and the key message
$w_k(m,l)\in\{1,2,\cdots,2^{(n_1R_s'(\hv))}\}$. These latter
messages will allow us to generate the key $K(m)$
of the super block $m$.

Since $b\rightarrow\infty$,
$a\rightarrow\infty$, $n_1\rightarrow\infty$, it can be
easily shown that the rates $R_o(\hv)$, $R_s'(\hv)$,
and $R_s''(\hv)$ are achievable within a given block.
Furthermore, the average key rate, $\mathbb{E}[R_s'(\hv)]$,
is achievable within any super-block (see, e.g.,~\cite{Gopala:On08}).

We finally argue that the equivocation rate at the
eavesdropper can be made arbitrarily close to the message rate
with the proposed scheme. Here, we consider equivocation
computation per block, which will imply the equivocation
computation for the overall message. For a given block
$(m,l)$, the security of $w_1(m,l)$ follows from the one-time
pad encryption (as the key is secured from the
eavesdropper~\cite{Shannon:Communication49}) and the security
of $w_2(m,l)$ follows from the wire-tap channel result along with the
secure rate choice $R_s'(\hv)$ and $R_s''(\hv)$~\cite{Wyner:The75}.
We note that during the first super-block $w_1(1,l)$ is not
encrypted. However, this will not affect the overall secrecy
of the data as $b\to\infty$. Hence, the equivocation rate
can be made close to the message rate as $b\rightarrow\infty$,
$a\rightarrow\infty$, and $n_1\rightarrow\infty$.

The achievable rate is then minimized over $\hv$ to
satisfy the delay limitation
and then maximized over all power control policies and
functions $q(\hv)$ (used to allocate rate for $w_2$).
This proves the desired result.
\end{proof}

The final step in this section is to prove the asymptotic
optimality of the proposed security scheme in the high
SNR regime. The following result establishes this objective
by showing that the upper and lower bounds of
Theorems~\ref{thm:upperFullCSI} and~\ref{thm:lowerFullCSI}
match in this asymptotic scenario.

\begin{lemma} \label{thm:lemmafull}

In an asymptotic regime of high SNR,
i.e., $\bar{P}\to\infty$, the delay limited secrecy capacity
is given by
\begin{eqnarray}
\lim\limits_{\bar{P} \to \infty} C_{s,d}^{(F)}
&=& \mathbb{E}_{h_m>h_e}
\left[
\log \left( \frac{h_m}{h_e} \right)
\right], \label{eq:Lemma3eq1}
\end{eqnarray}
assuming that ${\mathbb E}\left[\frac{1}{\min(h_e,h_m)}\right]$
is finite. Moreover, the capacity is achieved by the proposed
one-time pad encryption scheme coupled with the key renewal process.
\end{lemma}

\begin{proof}
We only need to consider the lower bound since the right hand side
of \eqref{eq:Lemma3eq1} is the ergodic secrecy capacity in
the high SNR regime, which is by definition an upper bound
on the delay limited secrecy capacity. To this end, we set
$q(\hv) = h_e$ resulting in $R_s''(\hv)=0$. Furthermore, we
let $P(\hv)=\frac{c}{\min(h_e,h_m)}$, where $c$ is a constant,
which is chosen according to the average power constraint.
The achievable rate expression in the high SNR regime is
then given by
\begin{equation}
\lim\limits_{\bar{P} \to \infty}
R_{s,d}^{(F)} = \lim\limits_{\bar{P} \to \infty}
\min \limits_{\hv}
R_o(\hv),
\end{equation}
where $R_o(\hv)$ is chosen to satisfy
\begin{eqnarray}\label{eq:Lemma3eq2}
\mathbb{E}[R_o(\hv)] &\leq& \mathbb{E}[\log(1+P(\hv)h_m)-
\log(1+P(\hv)h_e)]^+]\nonumber\\
R_o(\hv) &\leq& \log(1+c)
\end{eqnarray}

As $\bar{P} \to \infty$, it is easy to see that $c\to \infty$
since ${\mathbb E}\left[\frac{1}{\min(h_e,h_m)}\right]$ is
finite, implying that the second constraint in \eqref{eq:Lemma3eq2}
is loose. Also, it is easy to see that the first constraint
converges to the right hand side of the lemma. Then, by choosing
$R_o(\hv)=\mathbb{E}_{h_m>h_e}
\left[
\log \left( \frac{h_m}{h_e} \right)
\right]$, both constraints of \eqref{eq:Lemma3eq2}
are satisfied and our result is proved.
\end{proof}

The above claim is validated numerically in Fig.~$2$, where
Chi-square distribution of degree $n=4$ is used
for the statistics of channel gains of the legitimate
receiver and the eavesdropper (the gains are assumed
to be independent). In our simulation, we set $q(\hv)=h_e$
(hence $R_s''=0$) and use channel inversion power control
policy for the achievable rate.
Remarkably, even with the suboptimal choice of
$q(\hv)$ and $P(\hv)$, lower and upper bounds coincides
at the high SNR regime.

\begin{figure}[t] 
    \centering
    \includegraphics[width=1\columnwidth]{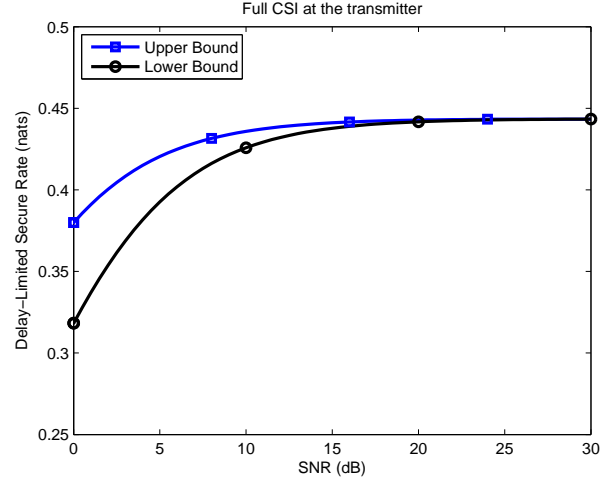}
    \caption{Simulation results for the full CSI case}
\end{figure}


\subsection{Only Main Channel CSI Scenario}

In this section we assume that only the legitimate
receiver CSI is available at the transmitter. First, we have
the following upper bound.

\begin{theorem}\label{thm:upperMainCSI}

The delay-limited secrecy capacity when only the
legitimate receiver channel state is available at
the transmitter, $C_{s,d}^{(M)}$,  is upper bounded by
\begin{eqnarray}
C_{s,d}^{(M)} \leq
\max \limits_{{}^{\quad \: P(h_m)}_{\textrm{s.t. }
{\mathbb E} [P(h_m)]\leq \bar{P} }}
\min
\left\{  R_s^{(M)} ,  R_d^{(M)}  \right\}
\end{eqnarray}
\noindent where $R_s^{(M)}$ and $R_d^{(M)}$ are given as follows.
\begin{eqnarray}
R_s^{(M)}&=& \mathbb{E} \left[\log(1+P(h_m)h_m)
- \log(1+P(h_m)h_e)\right]^+\nonumber\\
R_d^{(M)} &=& \min\limits_{h_m} \log(1 + P(h_m)h_m )\nonumber
\end{eqnarray}
\end{theorem}

\begin{proof}
The proof follows the same argument as that of
Theorem~\ref{thm:upperFullCSI} with the power
control policy $P(h_m)$.
\end{proof}

The achievability scheme in this scenario is different from the
previous scenario in two key aspects: 1) the lack of knowledge
about $h_e$ forces us to secure the whole delay sensitive traffic
with the one time pad approach (i.e., setting the rate of $w_2$ to
zero) to overcome the secrecy outage phenomenon and 2) the binning
scheme of the key renewal process must now operate on the level
of the super-block to average-out the fluctuations in $h_e$. On
the other hand, the delay sensitive packet must be decoded after
each block which makes it more challenging to use it as a
randomization for hiding the key. The achievable rate reported
in the following result is obtained by superimposing the binning
scheme used in~\cite{Gopala:On08}, to achieve the ergodic secrecy
rate for the key, on the delay limited traffic (secured by the
key bits sent in the previous super-blocks).

\begin{theorem}\label{thm:lowerMainCSI}

The delay-limited secrecy capacity in the
only main CSI scenario is lower bounded as follows.
\begin{eqnarray}
C_{s,d} ^{(M)} \geq R_{s,d}^{(M)} =
\max \limits_{{}^{\quad \: P(h_m)}_{\textrm{s.t. }
{\mathbb E} [P(h_m)]\leq \bar{P} }}
\min
\bigg\{
R_s, R_d^{(M)}
\bigg\},
\end{eqnarray}
where $R_s$ and $R_d^{(M)}$ are given as follows.
$$R_s = {\mathbb E}
[\log(1+P(h_m)h_m)-R_{s,d}^{(M)}
-\log(1+P(h_m)h_e)]^+$$
$$R_d^{(M)} = \min\limits_{h_m} \log(1 + P(h_m)h_m )$$
\end{theorem}

\begin{proof}[Sketch of the Proof] First, fix
a power control policy $P(h_m)$.
We then divide the channel uses into super-blocks and
further divide each super-block into blocks as
done in the proof of Theorem~\ref{thm:lowerFullCSI}.
In this scenario, we utilize the achievable secure rate
within a block only for the key generation. That is,
data is transmitted using only the one-time pad encryption
in contrast to the scheme used in Theorem~\ref{thm:lowerFullCSI}.
More specifically, the key is decoded at the end of each
super-block whereas the data packets are still decoded
block by block using the key sent in the previous super-block.

A given message $w\in\{1,2,\cdots, 2^{nR_{s,d}^{(M)}}\}$,
is divided into $ba$ data packets, each represented by $D(m,l)$,
where each packet is sent with rate $R_{s,d}^{(M)}$ during
the corresponding block of the channel. The data packet
$D(m,l)$ is transmitted along with the generated key using
the one-time pad scheme. Initial key generation and key renewal
is similar to the scheme in Theorem~\ref{thm:lowerFullCSI}.
For any super-block $m$, we use the key $K(m-1)$ for the
one-time pad, and also generate a new key $K(m)$ for the
use in the next super-block.

For any given block $(m,l)$, we use the $n_1R_{s,d}^{(M)}$
remaining bits from the key $K(m-1)$ and
denote corresponding bits as $\tilde{K}(m,l)$, where
we set $K(0)= \emptyset$.
These bits are used in a one-time pad scheme to construct
$D_o(m,l) = D(m,l) \oplus \tilde{K}(m,l)$. The encrypted
bits are then mapped to a message
$w(m,l)\in\{ 1,2,\cdots,2^{n_1R_{s,d}^{(M)}} \}$.
At this point, we choose the rate of this message to satisfy
$R_{s,d}^{(M)} \leq R_d^{(M)}$ to allow a fixed
rate transmission for every fading state, i.e., to
satisfy the delay limitation. For the key
renewal process, the binning scheme is constructed as in the
achievable scheme used in~\cite{Gopala:On08}. The output bits are
then grouped in blocks with rates given by
$\log(1+P(h_m)h_m)-R_{s,d}^{(M)}$.
We then combine those bits with the $R_{s,d}^{(M)}$ reserved for the
encrypted data packet and encode them using a capacity achieving
codebook (for the main channel). Each codeword is decoded at the
end of the block releasing the delay sensitive packet. In order
to decode the key bits, on the other hand, one must wait till
the end of the binning codeword (i.e., a super-block). Following
the argument given in~\cite{Gopala:On08}, one can see that
the following key rate can be achieved with perfect secrecy
(as $b\rightarrow\infty$, $a\rightarrow\infty$,
and $n_1\rightarrow\infty$).
$$ R_s= {\mathbb E}[
\log(1+P(h_m)h_m)-R_{s,d}^{(M)}
-\log(1+P(h_m)h_e)
]^+. $$
Therefore, if $R_{s,d}^{(M)}$ is chosen as in the theorem, there
will be enough key bits to encrypt the message of the following
block. Similar to Theorem~\ref{thm:lowerFullCSI},
the un-encrypted messages during the very first block
becomes negligible as $b\rightarrow\infty$, and the secrecy
requirement can be satisfied.
The achievable rate is then maximized over all
power control policies satisfying the average
power constraint to obtain the desired result.
\end{proof}

Interestingly, one can easily verify that for a wide class
of invertible channels (i.e.,
${\mathbb E}\left(\frac{1}{h_m}\right)$ is finite),
the rate $R_{s,d}^{(M)}$ is non-zero.
Numerical results are provided in Fig.~$3$, where
Chi-square distribution of degree $n=4$ is used for
the channel gains. Here, channel inversion power control
policy is used for both the upper and lower bounds.
The non-zero delay limited rate is evident
in the figure.

\begin{figure}[t] 
    \centering
    \includegraphics[width=1\columnwidth]{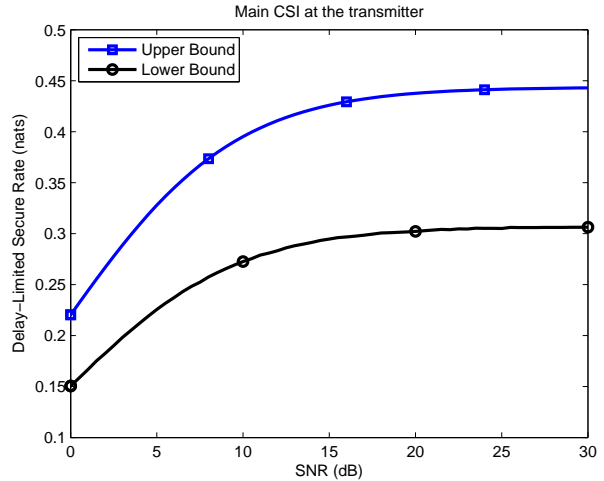}
    \caption{Simulation results for the main CSI case.}
    \label{OnlyMainSim}
\end{figure}


\section{Conclusion}

We have studied the delay-limited secrecy capacity of the slow-fading
channel under different assumptions on the CSI at the transmitter.
Our achievability arguments are based on a novel two-stage
scheme that allows for overcoming the secrecy outage phenomenon
for a wide class of channels. The scheme is based on sharing
{\em a delay tolerant} private key, using random binning, and then
using the key to encrypt the {\em the delay sensitive} packets in a one
time pad format. For the full CSI case, our scheme is further shown to
be asymptotically optimal, i.e., high SNR regime, for many relevant
channel distributions. When only the main channel CSI is available,
the two-stage scheme achieves a non-zero secure rate, under a strict
delay constraint, for invertible channels. Finally, one can easily
identify avenues for future works; three of them are immediate, namely
1) obtaining sharp capacity results for finite values of SNR, 2)
characterizing the optimal power control policies, and 3) extending
the framework to bursty traffic by allowing for buffer delays.




\end{document}